\documentclass[a4paper, 11pt,onecolumn,final,notitlepage,oneside]{article}
\usepackage{cmap} 
\usepackage[cp1251]{inputenc}
\usepackage[T1,T2A]{fontenc}
\usepackage[english,russian]{babel}
\usepackage{indentfirst}
\usepackage{geometry}
\usepackage{graphicx}
\usepackage[final]{epsfig}
\usepackage{multicol}
\usepackage{graphicx}
\usepackage{subfigure}
\usepackage{amsmath}
\usepackage{enumerate}
\usepackage{amssymb}
\usepackage{amscd}
\usepackage{hhline}
\usepackage{multirow}
\usepackage{setspace}
\usepackage{makeidx}
\usepackage{textcomp}
\usepackage{amsfonts}
\usepackage{afterpage}
\usepackage{longtable}
\usepackage{cite}
\usepackage{rawfonts}
\usepackage{array}
\usepackage{amsopn}
 \graphicspath{{pictures/}}

\DeclareGraphicsExtensions{.pdf,.png,.jpg,.eps}

\numberwithin{equation}{subsubsection}

\makeatletter
\def\@seccntformat#1{\csname the#1\endcsname.\ } 
\def\@biblabel#1{#1.} 
\makeatother
\addcontentsline{toc}{subsubsection}{References}
\addto\captionsrussian{}

\begin{document}

\title{\normalsize 
\begin{flushleft}
{subject:\textit{\,\,relativistic Kinematics}}
\end{flushleft}
\begin{flushleft}
{PACS: 03.30.+p}
\end{flushleft}
\vspace{\baselineskip}
 \normalsize \bf ON THE EQUATIONS OF THE INVERSE KINEMATICS PROBLEM}
\author{\bf \small V. V. Voytik\,\\
\small \itshape graduate student, Department of General and Teoretical Physics, \\ \small \itshape Bashkirian State Pedagogical University, \\ \small \itshape October Revolution st., 3a, Ufa, 450000, Russia\\
\small \itshape e-mail: voytik1@yandex.ru\\
\small \itshape Received 02.01.2014 г.\\
\small \itshape Submitted in STFI, 2014, no.1.\footnote{Due to a technical error, the originally published version of this paper has a large number of incorrect references. These errors have been corrected here.}\\}
\date{}
\maketitle
\renewcommand{\abstractname}{}

\begin{abstract}
The paper derived differential equations which solve the problem of restoration the motion parameters for a rigid reference frame  from the known proper acceleration and angular velocity of its origin as functions of proper time. These equations are based on the well-known transformation to an arbitrary rigid non-inertial reference frame, which generalized the Lorentz transformation, takes into account the fact of her proper of Thomas precession and rigid fixation  of the metric form reference frame. The role of this problem in physics is that all such reference frames found with the same characteristics will have the same properties. Another important advantage of these equations is that they allow you to find the motion of the non-inertial reference frame with respect to an arbitrary inertial frame. The resulting equations are nonlinear differential vector equation of the first order Riccati type
\[\mathbf{{\dot{v}}'}=\frac{d\,\mathbf{{v}'}}{dt}=\mathbf{{W}'}-(\mathbf{{W}'{v}'})\mathbf{{v}'}-\mathbf{\Omega}'\times \mathbf{{v}'}\]
and known system of 3 equations
\[\frac {1}{2}\,e^{\alpha\lambda\nu }a^{ \lambda\beta } \frac{da^{\alpha \beta } }{dt}=\omega'^{\nu }\,,\] which allows to determine the orientation of the coordinate system for a given function on the right side of the equation
\[ {\boldsymbol\omega }'=\mathbf{\Omega}'-\frac{1-\sqrt{1-{{{{v}'}}^{2}}}}{{{{{v}'}}^{2}}}\,\mathbf{{v}'}\times \mathbf{{W}'}\,,\] which are type angular velocity. The consequences of the obtained equations have been successfully verified on the example of a uniformly rotating disk.
\end{abstract}

\begin{flushleft}
{\bf{Introduction}}
\end{flushleft}

In addition to studying compound motions (that is, motions in two mutually moving frames of reference), kinematics as a science has 2 main tasks. If the velocity function $\mathbf{v}(t)$ is known, which determines the translational motion of the origin of the rigid non-inertial frame $s'$, and the self-rotation matrix $a^{\alpha\beta}(t)$ as a function of proper time $t$ , then its proper characteristics are also known, i.e. proper acceleration $\mathbf{W}'(t)$ and proper angular velocity $\mathbf{\Omega}'(t)$. This problem of determining the proper characteristics of an arbitrary reference frame from the known parameters of motion will be called the direct problem of kinematics. Then the inverse problem of kinematics is to find the parameters of the most general transformation into a group of all non-inertial frames of reference having a given proper acceleration and proper angular velocity. From this definition, the important role of this problem in physics is clear. All rigid reference frames found in this way with the same characteristics will have the same properties, i.e., for them
the generalized principle of relativity of A. A. Logunov will be fulfilled \cite[sect. 18, p. 126, p. 142 in Rus. ed.]{1}. If the proper acceleration and proper angular velocity of rigid reference frames are equal to zero, then such reference frames are inertial and the generalized principle of relativity turns into the principle of relativity of A. Einstein.

The usual solution of the inverse problem of kinematics in classical physics consists in the initial solution of the equation for the proper rotation matrix and substitution of the resulting solution into the equation that determines the proper acceleration (see below, respectively, the formulas \eqref{21} and \eqref{20}). In the relativistic case, such an approach to the inverse problem is incorrect, since it does not take into account the presence of Thomas's proper precession in a non-inertial reference frame moving forward. Meanwhile, Thomas's own precession at relativistic speeds cannot be neglected.

The equations of the inverse problem of kinematics proposed in this paper and the scheme for its solution are based on a remarkable transformation from a laboratory inertial frame of reference to an arbitrarily moving rigid non-inertial reference frame, which was finally proposed by Nelson \cite[equations (9a) and (9b)]{2}, \cite[equations (1),(2)]{3} \footnote{To avoid misunderstandings, it should be clarified that the author of this transformation in [3] implicitly considered the rotation defined by the matrix $a^{\alpha\beta}(t)$ to occur in the laboratory frame of reference. Here it is assumed that $a^{\alpha\beta}(t)$ is  matrix of proper rotation.}.
This holonomic general transformation automatically takes into account the facts of the existence of Thomas proper precession and the fixed metric form  of a rigid reference frame (eg \cite{4}).

The paper is structured as follows. Section \ref{f} shows how the direct problem of kinematics is solved in the relativistic case for an arbitrary rigid motion of a non-inertial reference frame. In Section \ref{s}, the equations of the inverse problem of kinematics are derived, and in Section \ref{t}, the procedure for its solution is proposed. In Section \ref{four}, these equations are checked for consistency and self-consistency using the example of a uniformly rotating rigid disk.

\subsubsection{Solution of the direct problem of kinematics for a rigid reference frame}\label{f}

We call a \textit{rigid reference frame} such a reference frame, which the metric tensor can depend on the physical time $t$ of the reference origin only through its proper acceleration $\mathbf{W}$ of the reference origin  or through its proper angular velocity $\boldsymbol{\Omega}$ as a function of time. So its metric is
\begin{equation}\label{1.1}
	g_{ik}=g_{ik}(\mathbf{W}(t),\boldsymbol{\Omega}(t),\mathbf{r})\,,
\end{equation}
   where $\mathbf{r}$ are the Cartesian coordinates of the non-inertial frame. Condition \eqref{1.1}, together with the condition of non-curvature of space-time and a number of other natural assumptions, uniquely fixes the form of the metric $g_{ik}$. For the first time, the form of the metric of a rigid accelerated reference system was discovered by C. M{\o}ller \cite[formula (17)]{5}. He discovered that quadratic interval in this reference frame is written in a simple form 
	\begin{equation} \label{1.2}
ds^{2} =(1+\mathbf{Wr})^{2}  \, dt^{2} -d\mathbf{r}^{2} ,
\end{equation}
	(hereinafter $c=1$).
What is the shape of the interval in the event that the frame of reference, in addition to its proper acceleration, also rotates? The simplest (but not rigorous) answer to this question can be obtained if we notice that the coordinate velocities  in a non-rotating frame of reference $\mathbf{v}_n$ and in a rotating frame of reference $\mathbf{v}_r$ are related by the relation
	\[\mathbf{v}_n=\mathbf{v}_r+\boldsymbol{\Omega}\times \mathbf{r}\,. \]
Multiplying this equation by $dt$, it is clear that in \eqref{1.2} it is necessary to make a change 
\begin{equation}\label{1.4}
	d\mathbf{r}\rightarrow d\mathbf{r}+ \boldsymbol{\Omega}\times \mathbf{r}\,\,dt\,. 
\end{equation}
Thus, the square of the interval of a rigid accelerated and rotating reference frame  should have the form \cite[p. 331, formula (13.71)]{6} (adjusted for the absence of space-time curvature in SRT)
\begin{equation} \label{1}
ds^{2} =\left[\, (1+\mathbf{Wr})^{2} -(\mathbf{\Omega \, \, \times r})^{2} \right]^{} \, dt^{2} -2(\mathbf{\Omega \times r})d\mathbf{r}\, dt-d\mathbf{r}^{2}\, .
\end{equation}
The Riemann-Christoffel curvature tensor, taken from the metric determined from \eqref{1}, is indeed equal to zero, so that the assumption \eqref{1.4} is justified.

The equatio \eqref{1} itself is an exact 4-dimensional definition of the rigidity properties of a non-inertial reference frame in the special theory of relativity. It is only necessary to make a remark concerning the understanding of the conventionality of the term "rigidity". The fact is that the spatial interval determined by the metric taken from \eqref{1} has a form 
	\[dl^2=d\mathbf{r}^2+\frac{\left[(\mathbf{\Omega \, \, \times r})\,d\mathbf{r}\right]^2}{(1+\mathbf{Wr})^{2} -(\mathbf{\Omega \, \, \times r})^{2}}\]
that depends on time. This shows that the so-called rigid frame of reference, when changing its proper kinematic characteristics, preserves distances only along the angular velocity vector and along the radius from it, and $\mathbf{r}$ is a constant radius vector from the origin to a given point. Tangential distances (that is, distances tangent to a circle around the angular velocity vector) are not preserved. Therefore, a rigid frame of reference is only \textit{radially rigid}. An absolute understanding of the body rigidity, as the preservation of distances between any two of its points, for arbitrary movement in the special theory of relativity is impossible.

The transition to the reference frame, the metric of which has the form \eqref{1}, takes place in two steps. At the first stage, the transition to the reference frame $s$ is carried out: it consists in the transformation \cite{2}
\begin{equation} \label{2}
     T=\frac{\mathbf{vr}}{\sqrt{1-v^{2} } } +\int _{0}^{t}\frac{dt}{\sqrt{1-v^{2} } },
\end{equation}
\begin{equation} \label{3}
 \mathbf{R=r}+\frac{1-\sqrt{1-v^{2} } }{v^{2} \sqrt{1-v^{2} } } \mathbf{(vr)v}+\int _{0}^{t}\frac{\mathbf{v}dt}{\sqrt{1-v^{2} } } .
\end{equation}
We call such a transformation a \textit{special Lorentz-M{\o}ller-Nelson transformation}. It links the laboratory inertial frame of reference $S:(T,\mathbf{R})$ and the non-inertial frame of reference $s:(t,\mathbf{r})$.
In the classical case, the segment between any two points of a non-inertial rigid reference frame, which moves forward, does not change and is transferred parallel to itself. Therefore, translational motion is completely characterized by only one vector  $\mathbf{v}(t)$. In relativistic physics, due to the Lorentz contraction, the Cartesian axes of the reference frame relative to the laboratory frame generally look oblique and coincide with the axes of such an inertial frame of reference, which instantaneously comoving $s$. This instantaneously comove inertial frame of reference is different for each moment of time. Due to this reason, the segment between two points of the frame $s$ in the process of its movement changes relative to the laboratory reference frame $S$. However, since the transformation \eqref{2}, \eqref{3} parameter, as in the classical case, is only one vector $\mathbf{v}(t)$, then for brevity, such an orbital motion of a non-inertial rigid system can continue to be called \textit{relativistically translational}. In this case, it turns out \cite{2} that the reference frame $s$, in addition to its proper acceleration equal to
\begin{equation} \label{4}
 \mathbf{W}=\frac{\mathbf{\dot{v}}}{\sqrt{1-v^{2} } } +\frac{1-\sqrt{1-v^{2} } }{v^{2} (1-v^{2} )} \,\mathbf{(\dot{v}v)v}\,,
\end{equation}
also has its proper rotation with an angular velocity
\begin{equation} \label{5}
\mathbf{\Omega} = \mathbf{\Omega}_{T}=\frac{1-\sqrt{1-v^{2} } }{v^{2} \sqrt{1-v^{2} } }\,\,\mathbf{v\times \dot{v}}\,,
\end{equation}
where $\mathbf{\dot{v}}={d\mathbf{v}}/{dt}\;$. This rotation depends on the nature of its orbital motion and is Thomas's proper precession. From \eqref{4} and \eqref{5} imply that
\begin{equation} \label{6}
\mathbf{\Omega} _{T} =\frac{1-\sqrt{1-v^{2} } }{v^{2} }\,\, \mathbf{v\times W}\,.
\end{equation}

At the second stage, a transition is made to the reference frame $s'$ rotating around $s$ according to the transformation \cite{3}
\begin{equation} \label{7}
r'^{\alpha } =a^{\alpha \beta } r^{\beta } \,.
\end{equation}
Here and below, the Greek indices range over the values 1,2,3, and summation is implied over twice occurring indices. Also, since the coordinate system is assumed to be rectangular, we will not distinguish between co- and contra- components of tensors. Given these conventions for the rotation matrix $a^{\alpha \beta}$, we can write the ratios of orthogonality 
\begin{equation} \label{8}
a^{\beta \alpha} a^{\gamma \alpha} =a^{\alpha \beta} a^{\alpha\gamma} =\delta ^{\beta\gamma}
\end{equation}
and equality of "annihilation"
\begin{equation} \label{9}
 e^{\alpha \mu \nu } a^{\mu \beta } a^{\nu \gamma } =e^{\mu \beta \gamma } a^{\alpha \mu } ,\,\,\,\,
 e^{\alpha \mu \nu } a^{\beta \mu } a^{\gamma \nu } =e^{\mu \beta \gamma } a^{\mu \alpha }\,,
 \end{equation}
where $e^{\alpha \beta\gamma}$ is a unit tensor perfectly antisymmetric in all indices, equal to 1 if the order of the indices is cyclic and -1 if this order is anticyclic. The name of the equalities \eqref{9} is determined by the fact that the combination of two rotation matrices on the left side of the equality is transformed into one rotation matrix on the right side \eqref{9}. It is easy to verify the validity of the equalities \eqref{9} for a specific representation of the rotation matrix, for example, in the coordinates: axis of rotation $\mathbf{n}$, angle of rotation $\mathbf{\phi}$ 
	\[a^{\alpha \beta}=\delta^{\alpha \beta}\cos\phi+n^{\alpha}n^{\beta}(1-\cos\phi)+e^{\alpha \beta\gamma}n^{\gamma}\,\sin\phi\,.\]
Reverse relation \eqref{7} is
\begin{equation} \label{10}
r^{\alpha } =a^{\beta \alpha} r'^{\beta }\,.
\end{equation}
Substituting it into \eqref{2}, \eqref{3} we get that the general transformation to an arbitrary rigid frame of reference looks like
\begin{equation} \label{11}
T=\frac{v^{\alpha } a^{\beta \alpha} r'^{\beta } }{\sqrt{1-v^{2} } } +\int _{0}^{t}\frac{dt'}{\sqrt{1-v^{2} } }
\end{equation}
\begin{equation} \label{12}
 R^{\alpha } =\left\{a^{ \beta \alpha} r'^{\beta } +\frac{1-\sqrt{1-v^{2} } }{v^{2} \sqrt{1-v^{2} } } v^{\alpha } v^{\gamma } a^{\beta \gamma} r'^{\beta } \right\}+\int _{0}^{t}\frac{v^{\alpha } dt'}{\sqrt{1-v^{2} } }
  \end{equation}
We call transformation \eqref{11}, \eqref{12} \textit{the general Lorentz-M{\o}ller-Nelson transformation}.

In the $s'$ frame, the change of an arbitrary vector $\mathbf{r'}$ associated with the reference frame $s$ obeys the equation
\[\frac{d\mathbf{r'}}{dt} =-\, \boldsymbol{\omega'} \times \mathbf{r'}.\]
 Here $\boldsymbol{\omega'}$ is the vector of the angular velocity of rotation $s'$ relative to the $s$ frame expressed in the $s'$ coordinate system. Substituting $r'^{\alpha}=a^{\alpha \beta}r^{\beta}$ here and assuming that $r^{\beta} =\texttt{const}$ we get that this matrix must satisfy the equation
\begin{equation} \label{14}
 \frac {da^{\alpha \beta}}{dt} =-e^{\alpha\mu\nu} \omega'^{\mu} a^{\nu \beta}\,.
 \end{equation}
Multiplying \eqref{14} by $e^{\alpha\lambda\nu}a^{\lambda\beta}/2$ we get
 \begin{equation} \label{15}
 \frac {1}{2}\,e^{\alpha\lambda\nu }a^{ \lambda\beta } \frac{da^{\alpha \beta } }{dt}=\omega'^{\nu }\,.
\end{equation}
Let us choose in \eqref{15} a specific representation of the rotation matrix $a^{\alpha \beta}$ in some system of 3 orientation angles. Then, if the vector $\boldsymbol{\omega'}$ is known as a function of time, then the system of 3 differential equations \eqref{15} is actually the problem of restoring the rotation angles from the known angular velocity. As first shown by Darboux \cite[ch. 3, p. 130, sect. 3.12 in Rus. ed.]{7} this problem is reduced to finding one particular solution of the Riccati-type equation.

If we substitute the transformation \eqref{10} into the interval \eqref{1} (where $\mathbf{\Omega=\Omega}_{T}$ ), then we get an interval of the same form \eqref{1}, but with new proper acceleration and new angular velocity equal to
\begin{equation} \label{16}
 W'^{\alpha } =a^{ \alpha \beta} W^{\beta }
 \end{equation}
\begin{equation} \label{17}
 \Omega '^{\alpha } = a^{ \alpha \beta}\Omega _{T}^{\beta } +\omega'^{\alpha }
\end{equation}
The total angular velocity of frame $s'$ is the sum of Thomas' proper precession of frame $s$, which comove  frame $s'$, and the rotation of frame $s'$ relative to $s$. Substituting the relations \eqref{4} and \eqref{5} into \eqref{16} and \eqref{17} we finally obtain the characteristics of the reference frame $s'$
\begin{equation} \label{18}
       {{{W}'}^{\alpha }}={{a}^{\alpha \beta }}\left[ \frac{1}{\sqrt{1-{{v}^{2}}}}\,{{{\dot{v}}}^{\beta }}+\frac{1-\sqrt{1-{{v}^{2}}}}{{{v}^{2}}(1-{{v}^{2}})}\,\,(\mathbf{v\dot{v}}){{v}^{\beta }} \right],
      \end{equation}
         \begin{equation} \label{19}
      {{{\Omega }'}^{\alpha }}={{a}^{\alpha \beta }}\,\frac{1-\sqrt{1-{{v}^{2}}}}{{{v}^{2}}\sqrt{1-{{v}^{2}}}}\,\,{{e}^{\,\beta \mu \nu }}{{v}^{\mu }}{{\dot{v}}^{\nu }}+{{{\omega }'}^{\alpha }}.
      \end{equation}
In classical physics, these equations have the form
\begin{equation} \label{20}
       W'^{\alpha }=a^{\alpha \beta }\dot{v}^{\beta }\,,
      \end{equation}
         \begin{equation} \label{21}
      {{{\Omega }'}^{\alpha }}={{{\omega }'}^{\alpha }}\,.
      \end{equation}
The second of these equations, taking into account \eqref{15}, can be solved for $a^{\alpha \beta}$ separately from the first one. Obviously, in the relativistic case, due to the presence of Thomas proper precession, the formula \eqref{19} is not independent of \eqref{18}.

\subsubsection{Differential equations for the inverse problem of relativistic kinematics}\label{s}

The system of 3  equations \eqref{18} and 3  equations \eqref{19} for the unknown 3 $\mathbf{v}$ components and 3 orientation angles, which define $a^{\alpha\beta}$, is complete and, hence, hase a solution. Before looking for it, let's pay attention to the following circumstance.
In both classical and relativistic cases, when solving kinematic problems, there is some mathematical inconvenience, which consists in the fact that the characteristics of a non-inertial frame of reference depend on the rotation matrix. This circumstance, although natural, somewhat complicates the mathematical writing of the equations for proper acceleration and angular velocity. Let us show that using some substitution it is possible to ensure that the equations for the inverse problem of kinematics will be written in a convenient vector form. After that, the order of its solution will be clear.

Let's introduce a new vector
   	 \begin{equation}\label{22}
		 {{v'}^{\beta }}={{a}^{\beta \alpha }}{{{v}}^{\alpha }}\,.
	 \end{equation}
     The physical meaning of the function $\mathbf{v'}(t) $ is that this value means the velocity of the origin of the frame $s$ - $\mathbf{v}(t) $, expressed in the coordinate system $s'$. From \eqref{22} it follows that
\begin{equation} \label{23}
 {{v}^{\alpha }}={{a}^{\beta \alpha }}{{{v}'}^{\beta }}.
 \end{equation}
Hence, due to the orthogonality of the rotation matrix
\begin{equation} \label{24}
                            {v}^2={v}'^{\,2}.
\end{equation}
Differentiating \eqref{24} we get that
\begin{equation} \label{25}
 \mathbf{v\dot{v}}=\mathbf{v'\dot{v'}}.
 \end{equation}
Let us now substitute \eqref{23}, \eqref{24} first into \eqref{18}. In this case, we will take into account \eqref{8}, as well as \eqref{25}. We have
\[ {{{W}'}^{\alpha}}={{a}^{\alpha \beta }}\left[\frac{1}{\sqrt{1-v'^2}}\,\frac{d(a^{\mu\beta}v'^{\mu})}{dt}+\frac{1-\sqrt{1-v'^2}}{v'^2(1-v'^2)}\,\,
          (\mathbf{v'}\dot{\mathbf{v'}})\, a^{\mu\beta} v'^{\mu}  \right]= \]
\[=a^{\alpha\beta}\cdot \frac{a^{\mu\beta}\dot{v}'^{\mu}+{{{\dot{a}}}^{\mu \beta }v'^{\mu}}}{\sqrt{1-{{{{v}'}}^{2}}}}+\frac{1-\sqrt{1-{{{{v}'}}^{2}}}}{{{{{v}'}}^{2}}(1-{{{{v}'}}^{2}})}\,\, \mathbf{(v'{\dot{v}}')}\, v'^{\alpha }\,.\]
Then we substitute the equations \eqref{14} into this equality. Then we get that
 \[{{{W}'}^{\alpha }}=\frac{\dot{v}'^{\alpha }+{{e}^{\alpha \lambda\mu }}{{{{\omega }'}}^{\lambda }}{{{{v}'}}^{\mu }}}{\sqrt{1-{{{{v}'}}^{2}}}}+\frac{1-\sqrt{1-{{{{v}'}}^{2}}}}{{{{{v}'}}^{2}}(1-{{{{v}'}}^{2}})}\,\,(\mathbf{{v}'{\dot{v}}'})\,{{{v}'}^{\alpha }}\,.\]
In vector form, this equation looks like
\begin{equation} \label{26}          \mathbf{{W}'}=\frac{{\mathbf{{\dot{v}}'}+{\boldsymbol{\omega} }'\times \mathbf{{v}'}}}{\sqrt{1-{{{{v}'}}^{2}}}}+\frac{1-\sqrt{1-{{{{v}'}}^{2}}}}{{{{{v}'}}^{2}}(1-{{{{v}'}}^{2}})}(\mathbf{{v}'{\dot{v}}'})\mathbf{{v}'}.
              \end{equation}
Similarly, in the equation \eqref{19} we obtain as a result of the same substitutions the following chain of calculations
\[{{{\Omega }'}^{\alpha }}={{a}^{\alpha \beta }}\frac{1-\sqrt{1-{{v}^{2}}}}{{{v}^{2}}\sqrt{1-{{v}^{2}}}}\,\,{{e}^{\beta \mu \nu }}{{v}^{\mu }}{{\dot{v}}^{\nu }}+{{{\omega }'}^{\alpha }}={{a}^{\alpha \beta }}\frac{1-\sqrt{1-{{v'}^{2}}}}{{{v'}^{2}}\sqrt{1-{{v'}^{2}}}}\,\,{{e}^{\beta \mu \nu }}{{a}^{\eta \mu }}{{{v}'}^{\eta }}\frac{d}{dt}({{a}^{\lambda \nu }}{{{v}'}^{\lambda }})+\]
	\[+{{{\omega }'}^{\alpha }}={{a}^{\alpha \beta }}\frac{1-\sqrt{1-{{{{v}'}}^{2}}}}{{{{{v}'}}^{2}}\sqrt{1-{{{{v}'}}^{2}}}}\,\,{{e}^{\beta \mu \nu }}{{a}^{\eta \mu }}{{{v}'}^{\eta }}  (a^{\lambda \nu }\dot{v}'^{\lambda }-e^{\lambda \theta \tau }{\omega '}^{\theta } a^{\tau \nu }v'^{\lambda })
+\omega'^{\alpha }=\]
\[=\frac{1-\sqrt{1-{{{{v}'}}^{2}}}}{{{{{v}'}}^{2}}\sqrt{1-{{{{v}'}}^{2}}}}\,\,({e}^{\nu \beta \mu } {a}^{\alpha \beta}  {a}^{\eta \mu }) {v}'^{\eta }  (a^{\lambda \nu }\dot{v}'^{\lambda }-e^{\lambda \theta \tau }{\omega '}^{\theta } a^{\tau \nu }v'^{\lambda })
+\omega'^{\alpha }.\]
In the first bracket we use the equality \eqref{9}. We get
\[\Omega '^{\alpha }=\frac{1-\sqrt{1-{{{{v}'}}^{2}}}}{{{{{v}'}}^{2}}\sqrt{1-{{{{v}'}}^{2}}}}\,\,e^{\beta \alpha \eta }a^{\beta \nu }v'^{\eta }(a^{\lambda \nu }\dot{v}'^{\lambda }-e^{\lambda \theta \tau }\omega '^{\theta }a^{\tau \nu }v'^{\lambda })+\omega '^{\alpha }=\]
   \[=\frac{1-\sqrt{1-{{{{v}'}}^{2}}}}{{{{{v}'}}^{2}}\sqrt{1-{{{{v}'}}^{2}}}}\left[ e^{\beta \alpha \eta }(a^{\beta \nu }a^{\lambda \nu })v'^{\eta }\dot{v}'^{\lambda }-e^{\beta \alpha \eta }{v}'^{\eta }{e}^{\lambda \theta \tau }{\omega }'^{\theta }(a^{\beta \nu }{a}^{\tau \nu }){v}'^{\lambda } \right]+{\omega }'^{\alpha }=\]
\[=\frac{1-\sqrt{1-{{{{v}'}}^{2}}}}{{{{{v}'}}^{2}}\sqrt{1-{{{{v}'}}^{2}}}}\left[ e^{\beta \alpha \eta }v'^{\eta }\dot{v}'^{\beta }-e^{\beta \alpha \eta }v'^{\eta }e^{\lambda \theta \beta }\omega'^{\theta }v'^{\lambda }\right]+{\omega }'^{\alpha }=\]
\[=\frac{1-\sqrt{1-{{{{v}'}}^{2}}}}{{{{{v}'}}^{2}}\sqrt{1-{{{{v}'}}^{2}}}}e^{\alpha \eta \beta }v'^{\eta }\left( \dot{v}'^{\beta }+e^{\beta \theta \lambda }{\omega }'^{\theta }v'^{\lambda }\right)+\omega '^{\alpha }\]
or in vector form
\begin{equation} \label{27}
            \mathbf{\Omega }'=\frac{1-\sqrt{1-{{{{v}'}}^{2}}}}{{{{{v}'}}^{2}}\sqrt{1-{{{{v}'}}^{2}}}}\,\,\mathbf{{v}'}\times (\mathbf{{\dot{v}}'}+{\boldsymbol\omega }'\times \mathbf{{v}'})+{\boldsymbol \omega }'.
                                    \end{equation}
The equation \eqref{27} is easy to solve with respect to $\boldsymbol{\omega'}$.
   We get
	\begin{equation} \label{28}
         {\boldsymbol\omega}'=\sqrt{1-{{{{v}'}}^{2}}}\,\,{\mathbf{\Omega} }'+\frac{1-\sqrt{1-{{{{v}'}}^{2}}}}{{{{{v}'}}^{2}}}\,({\mathbf{\Omega} }'\mathbf{{v}'})\mathbf{{v}'}-\frac{1-\sqrt{1-{{{{v}'}}^{2}}}}{{{{{v}'}}^{2}}}\,\,\mathbf{{v}'}\times \mathbf{{\dot{v}}'}.
                   \end{equation}
	Let's now substitute \eqref{28} into \eqref{26}. This leads to the following formula
	\begin{equation} \label{29}
        \mathbf{{W}'}=\mathbf{{\dot{v}}'}+\frac{(\mathbf{{v}'{\dot{v}}'})\mathbf{{v}'}}{1-{{{{v}'}}^{2}}}+{\mathbf{\Omega} }'\times \mathbf{{v}'}.
         \end{equation}
	Multiplying this equation by $\mathbf{{v}'}$ we get from here that
	\begin{equation} \label{30}
       \mathbf{v'\dot{v}'}=(1-{{{v}'}^{2}})\mathbf{{W}'{v}'}.    
			\end{equation}
	Let's now back-substitute \eqref{30} into \eqref{29}. We end up with a simple-looking equation
	\begin{equation} \label{31}
 \mathbf{{\dot{v}}'}=\frac{d\,\mathbf{{v}'}}{dt}=\mathbf{{W}'}-(\mathbf{{W}'{v}'})\mathbf{{v}'}-\mathbf{\Omega}'\times \mathbf{{v}'}.
      \end{equation}
	This equation is the vector Riccati equation for the unknown vector $\mathbf{v}'$. Substituting the equation \eqref{31} into \eqref{28} we obtain the second desired equation
\begin{equation} \label{32}
       {\boldsymbol\omega }'=\mathbf{\Omega}'-\frac{1-\sqrt{1-{{{{v}'}}^{2}}}}{{{{{v}'}}^{2}}}\,\mathbf{{v}'}\times \mathbf{{W}'}.
       \end{equation}
In fact, this equation is a relativistic refinement of the right side of equation \eqref{15}.
Thus, we see that in the formulas \eqref{31}, \eqref{32} for the eigencharacteristics of a non-inertial frame of reference there is no rotation matrix, so that they can be written in vector form. This greatly simplifies the formulas.

The equation \eqref{6} due to its vector nature in the reference frame $s'$ has the same form. It is only necessary to put strokes over all the quantities. Therefore, in the case when $\mathbf{\Omega=\Omega}_{T}$ the right side of the equality \eqref{32} vanishes and, consequently, $a^{\alpha\beta}=\texttt{const }$. Therefore, choosing a new coordinate system, we can achieve that $a^{\alpha\beta}=\delta^{\alpha\beta}$. This is how it should be, because when a non-inertial reference frame moves with a proper frequency equal to the Thomas precession frequency, its proper angular coordinates do not change.

If in the classical equations \eqref{20}, \eqref{21} it is similar to perform the substitution \eqref{23}, then we can obtain equations similar to \eqref{31} and \eqref{32}
\begin{equation} \label{33}
 \mathbf{{\dot{v}}'}=\frac{d\,\mathbf{{v}'}}{dt}=\mathbf{{W}'}-\mathbf{\Omega}'\times \mathbf{{v}'}\,,
      \end{equation}
     \begin{equation} \label{34}
       {\boldsymbol\omega }'=\mathbf{\Omega}'\,.
       \end{equation}
Comparison of the relativistic equations \eqref{31}, \eqref{32} with the classical ones \eqref{33}, \eqref{34} leads to the following obvious conclusions. First, \eqref{31}, \eqref{32} differ from \eqref{33}, \eqref{34} by the second members on the right side. These terms are relativistic corrections vanishing at $c\rightarrow \infty$ . In addition, \eqref{33}, \eqref{34} are independent equations, unlike \eqref{31} and \eqref{32}. In the relativistic case, the solution of the equation for the rotation matrix \eqref{32} depends on the solution \eqref{31}.

\subsubsection{Scheme for solving the inverse problem of kinematics.}\label{t}

Thus, the procedure for solving the inverse problem of kinematics is as follows. First, it is necessary to find the solution $\mathbf{v'}(t)$ of the nonlinear vector differential equation \eqref{31} of the first order. Substituting this solution into \eqref{32} reduces this equation to the problem of finding 3 rotation angles from the known angular velocity $a^{\alpha\beta}(t)$. Knowing these angles thus gives the matrix of proper rotation . Finally, substitution of already known values in \eqref{23} determines the parameter $\mathbf{v}(t)$ as a function of proper time and 6 constants, including 3 constants that determine the initial velocity and 3 initial orientation angles. Taking into account the three initial coordinates and the initial moment of time, the total number of initial constants that determine the law of motion of the reference frame $s'$ with given characteristics is 10, as it should be. The velocity function of the origin $s'$ $\mathbf{v}(t)$ can be represented in the usual form - in terms of laboratory time $T$, if we find the time $t$ as a function of $T$ from the equation \eqref{2}, where the change is made $\mathbf{r}= $0 
\[  \int\limits_{0}^{t}{\frac{dt}{\sqrt{1-{{{{v}'}}^{2}}(t)}}=T} \]
and substitute it into $\mathbf{v'}(t)$ and $a^{\alpha\beta}(t)$.

\subsubsection{Verification of equations on the example of a uniformly rotating disk}\label{four}

For the primary verification of the above equations, let us consider as an example a reference frame, the origin of which is located on the edge of a uniformly rotating hard disk, and the axes are associated with the material from which it is made. Such a frame of reference is accelerated and rotating, and the constant angular velocity of its rotation is perpendicular to
its constant proper acceleration $\mathbf{W'\Omega'}=0$. We will be interested in the proper characteristics of such a reference frame. It is easy to see that in this case the value
\begin{equation}\label{36}
  \mathbf{v}'=\frac{\mathbf{W'\times\Omega'}}{\Omega'^{\,2}}=\texttt{const}
\end{equation}
will be the solution
the first equation of the inverse problem of kinematics \eqref{31}.
From here
\begin{equation}\label{37}
  v'^{\,2}=\frac{W'^{\,2}}{\Omega'^{\,2}}\,.
\end{equation}
Note that when \eqref{36} is substituted into $\mathbf{v'\times W'}$, the value of this expression will be
\begin{equation}\label{38}
  \mathbf{v'\times W'}=\frac{W'^{\,2}}{{\Omega}'^{\,2}}\,\,\mathbf{\Omega'}=v'^{\,2}\mathbf{\Omega'}\,.
\end{equation}
Then substituting \eqref{38} into the second equation \eqref{32} we get that
\begin{equation} \label{39}
       {\boldsymbol\omega }'=\sqrt{1-v'^2}\,\,\mathbf{\Omega}'=\texttt{const}\,.
       \end{equation}
Although the solution \eqref{36} of the equation \eqref{31} is only a particular one, it is directly related to the correct dependence of the  velocity $\mathbf{v}(t)$ on time in \eqref{23} for the reference frame  origin. Indeed, in this case the matrix $a^{\beta\alpha}$ will depend on one rotation angle $\psi(t)$. Therefore, the projections of the velocity $\mathbf{v}(t)$ on the laboratory coordinate axes lying in the plane of rotation will periodically change according to the laws of sine and cosine, as it should be.
   The rotation angle $\psi(T)$ of the reference velocity vector $\mathbf{v}(t)$ is found by integrating \eqref{39} and taking into account the Einsteinian time dilation of the origin. We get
\[ \psi(T)=\int\sqrt{1-v'^2}\,\,{\Omega}'dt=\sqrt{1-v'^2}\,\,{\Omega}'t+\psi_0=\]
       \begin{equation}\label{40}
       =\sqrt{1-v'^2}\,\,{\Omega}'\cdot\sqrt{1-v'^2}\,\,T+\psi_0=(1-v'^2)\,\Omega'\, T+\psi_0\,.
     \end{equation}
On the other hand, this angle of rotation in the laboratory system is equal to
       \[\psi(T)=\Omega\, T+\psi_0\,,\]
where $\Omega$ is the angular velocity of disk rotation. Comparing this relation with \eqref{40} it is clear that
\begin{equation}\label{42}
   \Omega'=\frac{\Omega}{1-v^2}\,.
\end{equation}
Also using \eqref{37} and \eqref{24} it is clear that
\begin{equation}\label{43}
   W'=v\Omega'=\frac{v\Omega}{1-v^2}\,.
\end{equation}
The relations \eqref{42} and \eqref{43} have long been known (eg \cite[formulas for $\mathbf{\Omega}$ (42), (43)]{8}, \cite[formulas for $\mathbf{W}$ (30) and for $\mathbf{\Omega}$ (31)]{4}).

\begin{flushleft}
{\bf{Conclusion}}
\end{flushleft}
 The knowledge of one's proper acceleration and angular velocity as a function of one's proper time completely determines the entire class of physically equivalent reference frames. This problem is solved using the equations of the inverse problem of kinematics. In section \ref{s} of this paper, the equations of the inverse problem of kinematics were presented and in section \ref{four} were verified. They are a simple non-linear Riccati-type first-order vector differential equation \eqref{31} and a well-known system of 3 first-order equations for a Darboux-type problem \eqref{15}, where $\mathbf{\omega}'$ equals \eqref{32}. As is known \cite[ch. 3, sect. 3.12, p. 130 in Rus. ed.]{7}, the  system \eqref{15} can be reduced to finding one particular solution of the Riccati equation, so there may be some analogy between the space $\mathbf{v'}$ and the space of rotations.

The equations \eqref{31} and \eqref{32} were derived from the equations of the direct kinematics problem \eqref{18}, \eqref{19} which in turn are essentially based on the Lorentz-M{\o}ller-Nelson transformation. This circumstance can be considered as a disadvantage. Therefore, the next article will show another way to derive these equations \eqref{31}, \eqref{32}, without the need to refer to this transformation.

Another direction of research is independent subsequent verification of the presented method for solving the inverse kinematics problem. Such a check is possible when the characteristics of the reference system are such that these equations are solved. The simplest and most interesting is the case of constant acceleration, which will be considered in another article. In the general case, the solution of these differential equations has a complex analytical form, if any. However, with the development of computer methods and technologies, even the numerical solution of the inverse problem of kinematics is undoubtedly of physical interest.

\begin{flushleft}
{\bf{Gratitude}}
\end{flushleft}

The author is very grateful to Professor N. G. Migranov for useful discussions and support.

\textbf{After paper.} An equations (1.95), (1.131), which very similar to \eqref{31} was first established by Yaakov Friedman with the assistance of Tzvi Scarr in the book

9. Friedman, Y.; Scarr, T. \textit{Physical Applications of Homogeneous Balls.} Progress in Mathematical Physics, vol. 40, Springer Science+Business Media, New York, 2005, LLC, https://doi.org/10.1007/978-0-8176-8208-8

\end {document}